\documentclass[usenatbib]{mn2e}
\usepackage{epsfig}
\usepackage{amsmath}
\usepackage{natbib}
\usepackage{times}
\usepackage{bm}
\usepackage{color}

\newcommand{\kms}{\>{\rm km}\,{\rm s}^{-1}}

\newcommand{\solarmass}{\rm M_\odot}

\newcommand{\ms}{h^{-1} {\rm M_{\odot}}}
\newcommand{\mpch}{h^{-1} {\rm Mpc}}
\newcommand{\kpch}{h^{-1} {\rm kpc}}

\def\gsim { \lower .75ex \hbox{$\sim$} \llap{\raise .27ex \hbox{$>$}}}
\def\lsim { \lower .75ex \hbox{$\sim$} \llap{\raise .27ex \hbox{$<$}}}
\title{The Local Void: For or Against $\Lambda$CDM?}
\author[Xie et al.]
       {Lizhi Xie$^{1}$ \thanks{Email:lzxie@bao.ac.cn}, 
        Liang Gao$^{1,2}$,
        Qi Guo$^{1,2}$
\\
$^1$The Partner Group of Max Planck Institute for Astrophysics,
National Astronomical Observatories, Chinese Academy of Sciences, Beijing, 100012, China\\
$^2$Institute of Computational Cosmology, Department of Physics,
University of Durham, Science Laboratories, South Road, Durham DH1
3LE \\
}

\begin{document}

\maketitle
\label{firstpage}

\begin{abstract}
The emptiness of the Local Void has been put forward as a serious
challenge to the current standard paradigm of structure formation in
$\Lambda$CDM. We use a high resolution cosmological N-body
simulation, the Millennium-II run, combined with a sophisticated
semi-analytic galaxy formation model, to explore statistically 
whether the local void is allowed within our current knowledge of galaxy
formation in $\Lambda$CDM. We find that about $14$ percent of the
Local Group analogue  systems ($11$ of $77$) in our simulation are
associated with nearby low density regions having size and 'emptiness'
similar to those of the observed Local Void. This suggests that,
rather than a crisis of the $\Lambda$CDM, the emptiness of the Local
Void is indeed a success of the standard $\Lambda$CDM theory. The
paucity of faint galaxies in such voids results from a combination of
two factors: a lower amplitude of the halo mass function in the voids
than in the field,  and a lower galaxy formation efficiency in the
void haloes due to halo assembly bias effects. While the former is the
dominated factor, the later also plays a sizable role.  The halo
assembly bias effect results in a stellar mass fraction $25$ percent
lower for void galaxies when compared to  field galaxies with the same
halo mass. 
\end{abstract}


\section{Introduction}
\label{sec:intro}
Full sky galaxy surveys in the local Universe reveal the  striking
fact that a very large region around the Local Group is devoid of
galaxies. The feature was firstly noticed by \cite{tully88}. This Local Void
occupies a large fraction of the Local Volume (defined as a
sphere ${\rm 1 Mpc< D <8 Mpc}$ around the Milky Way).
 While this region is close enough that the
observational completeness limit is remarkably faint, still very few
galaxies been found in the Local Void, even with the most up-to-data
optical and HI surveys \citep{karachentsev04,karachentsev13}.

\cite{peebles10} claimed that the emptiness of the Local Void may be a
big challenge to the standard hierarchical structure formation
theory. These authors compiled a galaxy catalog with $562$ nearby
galaxies and found only $3$ galaxies contained in a region as big as
one third of the Local Volume.  Using an argument analogues
to the halo occupation model (HOD) of \cite{tinker09} (which assumes a
tight halo mass - galaxy luminosity relation) and the fact that the
amplitude of the halo mass function of the Local Void is one tenth of
the Local Volume \citep{gottlober03}, they claimed $19$ galaxies
should have been found in the Local Void, many more than those
detected in the real Universe. However, the argument of
\cite{peebles10} can be biased by two factors. First the halo mass
function adopted for the local void is based on one particular
simulation of \cite{gottlober03}. More simulations are needed to
increase statistics. Moreover, dark matter haloes, in particular the
low mass ones, suffer from the assembly bias effect, due to which the
properties of dark matter haloes can vary significantly in corresponding to
different environments
(e.g. \cite{gao05,gao06,wechsler06,li13,lacerna11}). Galaxies in the local void
are usually very faint and are expected to reside in the very low mass
haloes for which the assembly bias is strong. Since galaxy
formation processes can be affected by the assembly history of
dark matter haloes, it is unclear to what extend the HOD
approach holds in this regime.

\cite{tikhonov09} addressed the same problem with a different
approach. Using a high resolution dark matter only simulation, 
they found that, above a given circular velocity threshold
$25\kms$, the number of dark matter haloes in their simulated voids
exceed the count of observed dwarf galaxies by one order of magnitude. In
this approach, the model is crucially based on a poorly justified
assumption that the peak circular velocity of dark matter is identical to the
rotational speed of neutral hydrogen in observed nearby dwarf
galaxies. 

In this paper, We make use of a large dark matter simulation of
 a standard cosmology, combining it with the sophisticated semi-analytic
galaxy formation model of \cite{guo11}, to explicitly examine
whether the Local Void is allowed with our current understanding of
galaxy formation, or whether new physics is required to explain this
observation.  Note, there are a number of theoretical and
observational studies on Cosmic
Voids\citep[e.g.][]{mathis02,alpaslan14,pan12,kathryn11,sutter13,tavasoli13},
the Voids addressed in these studies are much larger on scale, and
their member galaxies are much brighter than what we study here.

The outline of our paper is as follows. We briefly introduce the
simulation and the semi-analytical galaxy formation model, as well as
the nearby galaxy catalog used for our study, in section 2. In section 3,
we  present our results. Finally we give a short summary and
discussion. 

\begin{figure*}
\hspace{0.13cm}\resizebox{8cm}{!}{\includegraphics{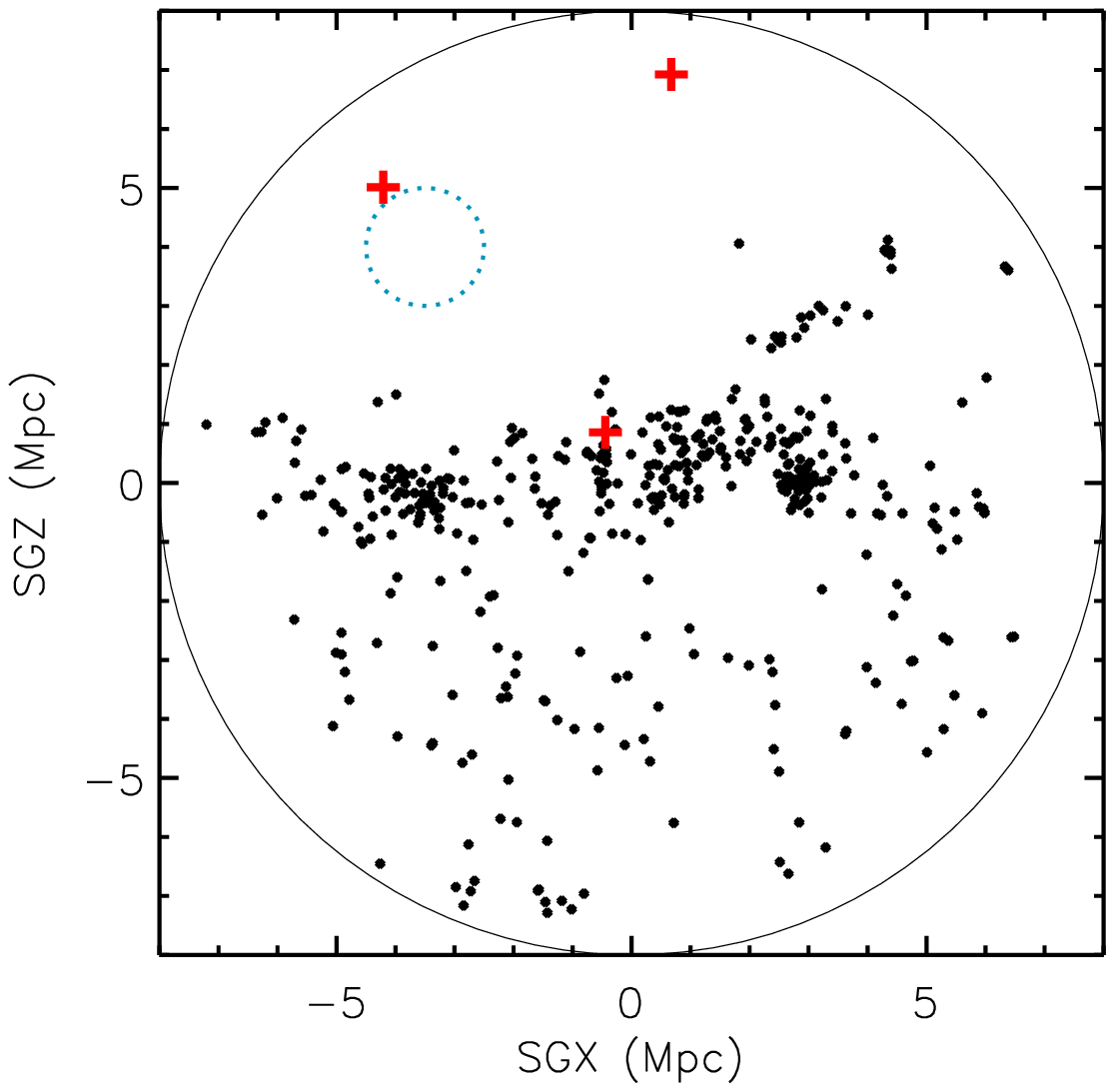}}
\hspace{0.13cm}\resizebox{8cm}{!}{\includegraphics{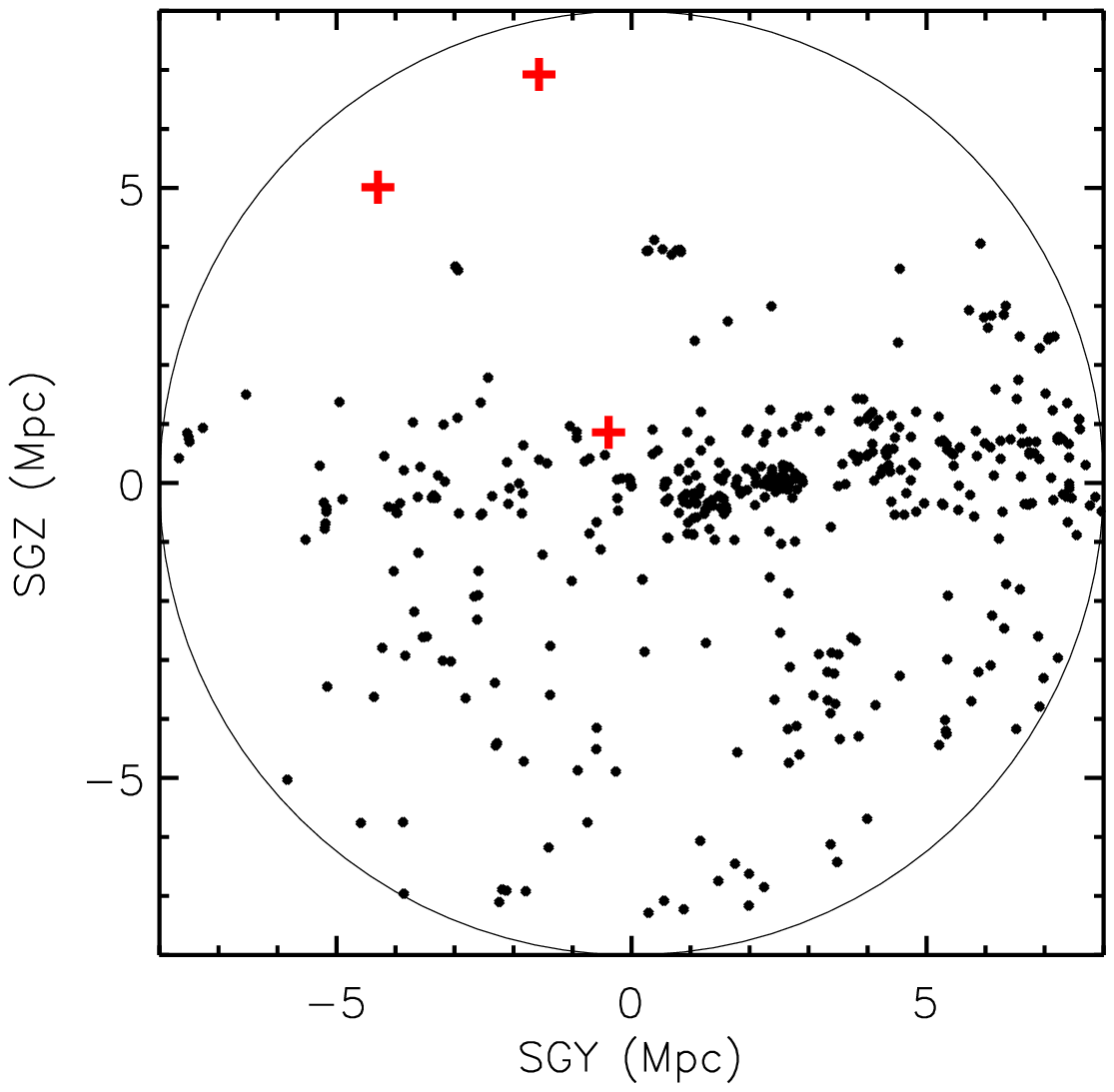}}
\caption{Distribution of nearby galaxies in the catalog of
  \protect\cite{karachentsev13}. Galaxies brighter than $m_B < 17.5$
  and located at distances ${\rm 1 Mpc < D < 8 Mpc}$ are shown in two
  orthogonal projections in super-Galactic coordinates. The red crosses
  are galaxies physically within the local void (see text for
  definition)  in 3 dimensions. The dashed blue circle bounds the
  projected distribution of the $3$ galaxies in the catalog of 
  \protect\cite{peebles10} but none of these are in the catalog used 
  here, because of the updated distance
  estimations(see the text for details).} 
\label{fig:lg_obs}
\end{figure*}

\section{The Simulation, the Semi-Analytic model and the Local Volume galaxy catalog}
We use a $10$-billion particle dark matter only simulation,
Millennium II (MSII), with a 100$\mpch$ cubic
volume. The simulation was run with the P-Gadget3
code \citep{springel05} with a mass resolution of $6.89\times 10^6 \ms$
and a force softening of $\epsilon=1 \kpch$, where $h = 0.73$. The
cosmological parameters are assumed to be $\Omega_m = 0.25$,
$\Omega_{\Lambda}=075$, and H = $h[100 h{\rm km  s^{-1}Mpc^{-1}}]$)
and the normalization $\sigma_8=0.9$. These values deviate somewhat
from the latest CMB results \citep{komatsu10,plank}. The small offset
is of no significant consequence for the topic discussed here as
galaxy formation is not very sensitive to the assumed cosmological
parameters \citep{wang08,guo13}.  

Dark matter haloes in our simulations are identified with a standard
friends-of-friends (FOF) group finder with a linking length 0.2 times
the mean inter-particle separation \citep{davis85}. For each FOF group,
we further identify self-bound and local over-dense subhaloes
using SUBFIND \citep{springel01}. All subhaloes containing more
than 20 particles are identified. The numerical data are stored at 68
times spaced roughly logarithmically and merger trees are
computed to follow the formation and merger history of each
halo/subhalo.

We populate dark matter haloes and subhaloes with galaxies using the
semi-analytic galaxy formation model recently developed by
\cite{guo11}. This model successfully reproduced 
various observed galaxy properties, in particular the faint end 
of the galaxy stellar mass function. This allows us to explore
the formation of very faint galaxies similar to those
in the Local Volume. 

We use the most recent nearby galaxy catalog by
\cite{karachentsev13}. It consists of several optical and HI
blind surveys, includes SDSS \citep{abazajian09} and the ${\rm H_{I}}$Parkes
All Sky Survey (HIPASS) \citep{wong06}. The catalog of 
\cite{karachentsev13} is complete at
70\% level for an apparent magnitude cut at $m_B<17.5$. This new
catalog contains $486$ galaxies brighter than $m_B<17.5$
within a region ${\rm 1 Mpc < D < 8 Mpc}$, which is significantly larger than the
previous catalog of \cite{karachentsev04}. The latter was adopted by the study
of \cite{peebles10}. In \cite{peebles10}, in addition to that sample,
the authors added $172$ more galaxies discovered by SDSS
survey and $53$ by HIPASS HI blind survey. 
The distances of these extra galaxies were estimated
with less secure radial velocities. A number of galaxies 
used in \cite{peebles10} are not included in the
catalog of \cite{karachentsev13}. This is because that SDSS and HIPASS
contain many spurious objects like stars or high velocity clouds,
which have been carefully excluded by \cite{karachentsev13}. 

The distribution of nearby galaxies in \cite{karachentsev13} is
displayed in Figure~\ref{fig:lg_obs} in two different projections
identical to those in figure 1 of \cite{peebles10}. Here we only show
galaxies brighter than $m_B<17.5$ and with distance ${\rm 1
  Mpc < D < 8 Mpc}$ from the Milky Way.  We illustrate
galaxies located in the 3-dimensional Local Void (see next section)
with red crosses. Some galaxies only appear in the void due to
projection. A similar plot was shown in
\cite{karachentsev13}. Compared to the figure 1 of
\cite{peebles10}, the $3$ galaxies they found in SGZ-SGX plane 
(the area enclosed by blue dashed circle in the left panel) are 
absent here. Instead, a few galaxies are
distributed in this region in a more diffuse way in projection.
The $3$ galaxies in \cite{peebles10} are excluded or found in different positions 
by the catalog of \cite{karachentsev13}. Note, the galaxy located in the
circled area of \cite{karachentsev04} is also removed in the updated
catalog. 

\section{Results}
\label{sec:lv}
\subsection{Identification of The Local Void from the nearby galaxy catalog}
The precise extent of the Local Void is subject to its particular
definition. \cite{peebles10} estimated its volume as about one third 
of the Local Volume. With the new, cleaner and more complete galaxy
uncatalogued of \cite{karachentsev13},  we define the Local Void with a
simple procedure, as follows, in order to make direct comparisons
with our numerical simulation. 

Motivated by the fact that only a few galaxies are sparsely distributed in
a large fan-shaped area of the Local Volume, we try to
delineate the Local Void with a conical geometry. We expect to
 find a large cone in the Local Volume containing only a
 few galaxies.  In practice, we scan over the Local Volume centered
 at the Milky Way, with a cone of solid angle $\Omega = \pi$, taking
 steps of $\frac{\pi}{20}$ along the spherical coordinates $\phi$ and
 $\theta$. At each step, we count the number of galaxies brighter
 than an apparent magnitude cut $m_B < 17.5$  within the truncated
 cone(a cone with apex cut off) and within a radial range 
${\rm 1 Mpc < D < 8 Mpc}$(the cone has spherical top and spherical bottom). We find
 the minimum number galaxies in a truncated cone with these conditions
 is $3$,  KK 246, AlFA ZOA J1952+1428 and Sag dIr. These $3$
galaxies have radial distances from the Local Group center $7.83$, $7.13$
and $1.04$ Mpc, respectively. It is certainly arbitrary to delineate
 the Local Void with a cone  with the solid angle $\Omega=\pi$. We
 further increase slightly the solid angle to  $\frac{4\pi}{3}$, and
 find the minimum number of galaxies in the new truncated cone
 increases by a factor $8$,  suggesting that the angle $\Omega = \pi$
 adopted here is a good empirical representation of the LV. Although
 it might be overly simplistic to describe the Local Void with a
 regular cone, we think it has its own advantage to set up a fair
 and direct comparison between the theory and the observations, as long
 as we apply the same procedure both to the data and to our numerical
 simulation.

\subsection{LV candidates selection from the numerical simulation}
\label{subsec:void}

\begin{figure}
 \includegraphics[width=0.45\textwidth]{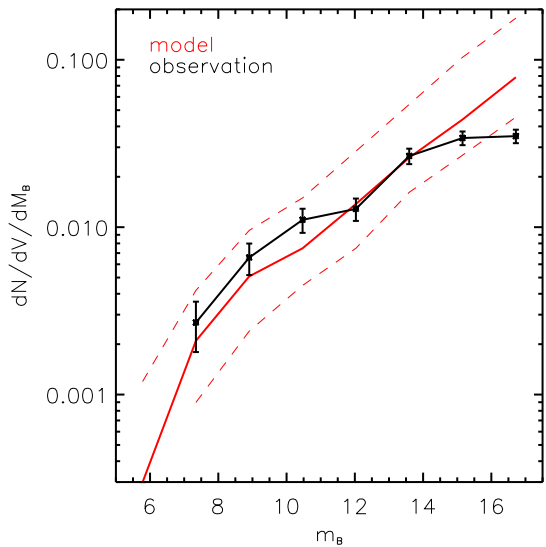}
 \caption{Abundance of galaxies as a function apparent magnitude in
   the observed Local Volume and in the simulated local volume
   samples. The solid black line shows results for the observational
   data with Poisson noise. The red solid line shows the median galaxy
   luminosity of the $77$ simulated local volume samples. The dashed
   lines show the $1 \sigma$ deviations.} 
\label{fig:lvlf}
\end{figure}

The Local Group is comprised of a pair of giant galaxies separated by a
distance of $0.77$Mpc, our Milky Way and M31. The Milky Way is a
spiral galaxy with a stellar mass about
$6.4\times10^{10}\solarmass$\citep{mcmillan11}, while M31 has slightly 
less massive halo but contains more stars $(10-15)\times 10^{10} \solarmass$
\citep{tuvikene12}. No massive galaxy clusters has been found
within $10$Mpc of the Milky Way.  To make fair comparisons 
between the simulation and the observations, we first identify 
local volume samples from our simulation with criteria as follows.

We first select Milky Way (MW) candidates from the simulated
galaxy uncatalogued that satisfy the constraints of being disk dominated,
 with a bulge to total mass ratio ${\rm M_{bulge}/M_{*} < 0.5}$, and
having a stellar mass similar to that of the Milky Way,
$5.4\times10^{10}\solarmass<M_{*}<7.4\times10^{10}\solarmass$. We 
further select local group samples by requiring a companion giant
galaxy with a stellar mass similar to that of the M31 $0.5\times
M_{\rm MW}<M_{*}<2 \times M_{\rm MW}$, within $1$Mpc from the 'Milky
Way'. Because the Milky Way and M31 are the brightest galaxies
within $1$Mpc, we also require that no bright
galaxies are present in the same volume besides the 'MW' and
'M31'. Finally we exclude systems with clusters ($10^{14}
\solarmass$) close by ($< 10 Mpc$). These constraints result 
in a sample of $77$ local groups in our full simulation box. 
\label{subsec:lf}

\begin{figure}
 \includegraphics[width=0.45\textwidth]{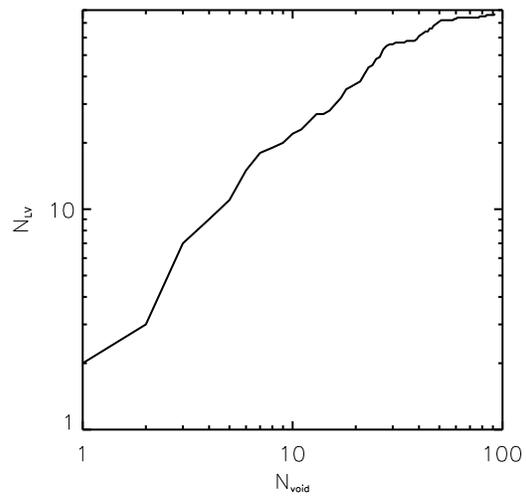}
 \caption{The cumulative number of the simulated local void samples
 ${\rm N_{LV}}$ as a function of the number of galaxy enclosed ${\rm N_{void}}$.}
 \label{fig:numd}
\end{figure}

\begin{figure*}
\centering
 \includegraphics[width=1.0\textwidth]{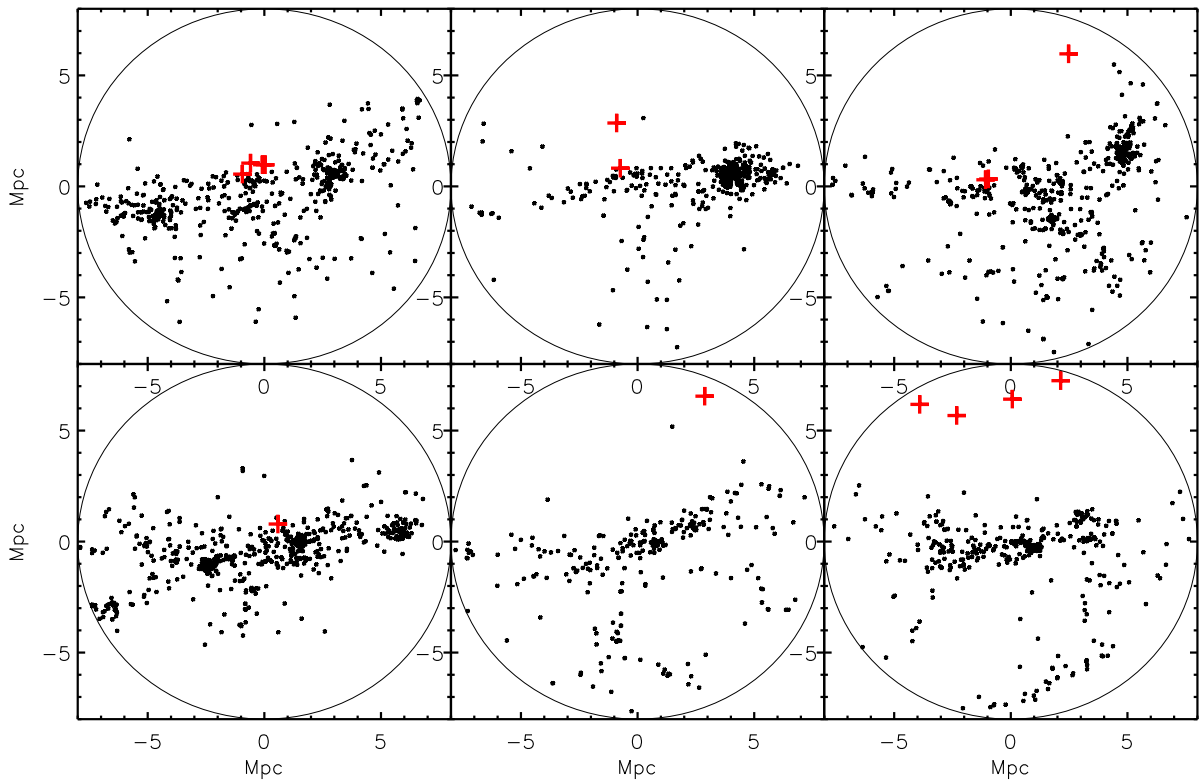}
\caption{Galaxies distribution in $6$ of our $77$ simulated 
  local volumes. We have selected samples which have
  the lowest abundance of galaxies in their local voids. Only the
  galaxies  brighter than $m_B < 17.5$ and located at distance
  ${\rm 1 Mpc < D < 8 Mpc}$ from their local volume centers are shown(black dots).
  The red crosses show galaxies physically within their local voids.}
\label{fig:void} 
\end{figure*}

It is important to see whether the luminosity function (LF) of the 
simulated local volumes agree with that of observed galaxies
~\citep{karachentsev13}. In Figure~\ref{fig:lvlf}, we show galaxy
number counts as a function of apparent magnitude for both the
observed Local Volume and our simulated local volume samples. Here we
use distance between galaxies and the MW to calculate the
 apparent magnitude. The black line with Poisson error bars is from
the observational data of \cite{karachentsev13}, the red solid line
shows the median values of our simulation. The dashed lines denote
$1 \sigma$ deviations. Clearly our simulations are in excellent
agreement with current observational data. No completeness correction
is applied here. Note that the galaxy catalog of \cite{karachentsev13}
was estimated to complete about $70-90\%$  at apparent magnitude
$m_{B}=17.5$, which corresponds to the flat faint end of the observed
 galaxy counts. Predictions from our galaxy formation model rather prefer
an continued increase in the galaxy number counts at the faint
end. This should be further constrained by future surveys extending
to fainter limits. 

We apply an identical procedure for identifying the Local Void in the
Local Volume to each of our simulated local volumes, i.e. finding the most
empty truncated cone shaped region. For each of them, we record the
corresponding galaxy number ${\rm N_{void}}$. We use the exactly
the same apparent magnitude cut $m_B=17.5$ to count the galaxies as
for the observational data. In Figure~\ref{fig:numd}, we plot the
cumulative number of local volume samples as a function of
${\rm N_{void}}$. The most empty void in our simulated samples only
contains $1$  galaxy, even emptier than the real Local
Void. When taking into account Poisson noise ($<=5$ galaxies),
there are in total $11$ local volume samples with a void as empty as
the real Local Void, more than $14\%$, suggesting that the 
emptiness of the local void is not uncommon.

It is interesting to visualize our local void
samples. In Figure~\ref{fig:void} we present distribution of 
galaxies within a distance ${\rm 1Mpc < D <8 Mpc}$ for $6$ of our
local volume samples with an associated void containing fewer than $5$
galaxies. The red crosses shown in the figure are the galaxies
actually in the voids in 3-d rather than chance projection. It is 
interesting that most of these system indeed appear very similar
to the real Local Volume as shown in the Figure~\ref{fig:lg_obs}.

\subsection{Void galaxy properties}
We explore in this section why the simulated local voids are short of
galaxies. One straightforward interpretation is that the halo
abundance is lower in under-dense regions as argued by
\cite{tinker09}. Below we examine whether this can fully account for the
emptiness of the local voids, or whether some environmental dependence of
galaxy formation can also play an important role.

\begin{figure}
\includegraphics[width=0.45\textwidth]{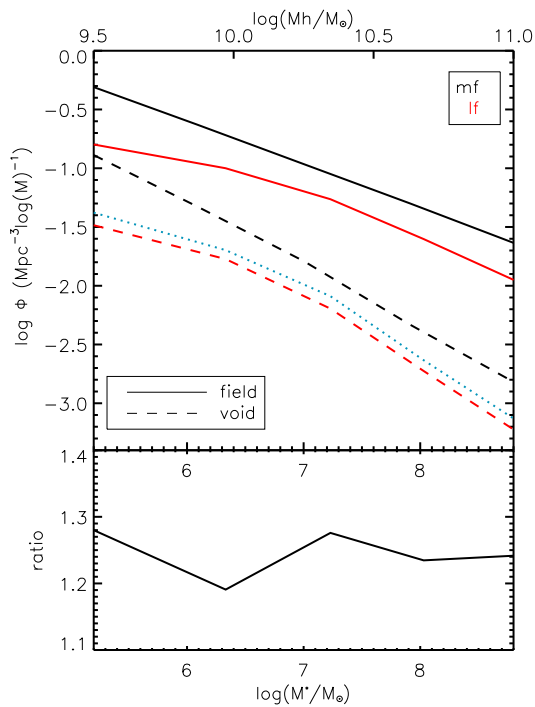}
\caption{Top panel: the abundance of haloes in the field 
(black solid lines) and in the voids(black dashed lines) as a function
  of halo mass (upper axis); the abundance of galaxies in the field
  (black dashed lines) and in the voids (red dashed line) as a function of
  stellar mass (lower axis). The blue dotted line show the predicted
  stellar mass function in the voids using a halo mass-to-stellar mass
  relation obtained for the field. Bottom  panel: the ratio between
  the blue dotted line and the red dashed line.}
\label{fig:mflf}
\end{figure}

In the top panel of Figure~\ref{fig:mflf}, we show the abundance of
dark haloes as a function of halo mass (upper axis) in the simulated
local voids (black dashed line). The abundance of galaxies as a function
of galaxy stellar mass (lower axis) in the simulated voids are shown
with red dashed line. Since the abundance of the dark haloes and 
galaxies in our local void samples is quite low, in order to increase
the statistics we stack the simulated local volume samples with the $20$
emptiest voids. The median values of $20$ samples are shown both for
the halo mass function and for the stellar mass function. For
comparison we also show the halo mass function and the stellar mass
function in the field(the whole simulating box) with black and red 
solid lines, respectively. It is remarkable to find that the different 
abundances of  haloes and galaxies are significant between the
field and the voids, especially at the high mass end, which excesses
an order of magnitude. The blue dotted lines show the predicted
stellar mass function if galaxy content of a dark matter halo only
depends on its halo mass. This is obtained by converting halo mass
in the voids to stellar mass using the halo mass to stellar mass
relation derived from the field. The extent to which it deviates from the
actual stellar mass function indicates the importance of environmental
effect on galaxy formation. In order to see the differences more
clearly, we plot the ratios between the blue dotted and the red dashed
line in the bottom panel. Clearly the dominate factor to account for
the emptiness of the voids in our simulation is the low abundance of
dark matter haloes, while the environmental dependence of galaxy
formation is relatively weaker but non-negligible, contributing 
at the level of $25$ percent.

\begin{figure}
\includegraphics[width=0.45\textwidth]{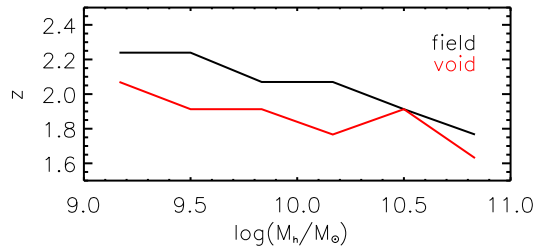}
\caption{The median halo formation time for haloes in the field (black)
  and in the voids (red) as a function of halo mass.}
\label{fig:ft}
\end{figure}

The different properties of galaxies in voids and those in the field 
result from the assembly bias  of dark matter haloes, namely
the formation history of a halo statistically depends on the
large scale environment. Assembly bias manifests itself in different
ways for different halo properties, for example halo formation time,
concentration parameter and spin parameter(e.g. Gao \&White
2006). As an example, we demonstrate the environmental dependence of 
halo formation time in Figure~\ref{fig:ft}. Here the halo formation
time is defined as the redshift at which a dark matter halo first
reaches half of its present day mass. In Figure~\ref{fig:ft}, we
display the median formation time of dark matter haloes
 as a function of halo mass for haloes in the voids(red) and those
 in the field(black). For given haloes' mass,  the
haloes in the voids typically formed later than those in the field
by d$z \sim 0.2$.

\section{Conclusion}
\label{sec:conclustion} 
We use a large dark matter only simulation MSII, combined
with the semi-analytic galaxy formation model of \cite{guo11,guo13}, to
study the faint galaxy population in the Local Volume with
luminosities similar to those in the full sky nearby galaxy catalog
compiled by \cite{karachentsev13}. In particular, we address the
issue of the emptiness of the Local Void, which has been put forward  as a
serious challenge to the standard structure formation theory by
\cite{peebles10} and \cite{tikhonov09}. Our results can be summarized
as follows. 

With the significantly updated nearby galaxy catalog
of \cite{karachentsev13}, we revise the 'emptiness' of the
Local Void. The updated catalog is a significant expansion over
previous ones used by \cite{peebles10}. These
galaxies are complete $70-80\%$ level for an apparent magnitude $17.5$  
and have secure distance estimations. We adopt a cone geometry with a
quarter of the Local Volume to define the Local Void. The most empty
region in the Local Volume only contains $3$ galaxies. When
applying an identical procedure to our simulated local volume
samples, we find about $14\%$ of the local volumes contain 
a low galaxy abundance region with size and the emptiness comparable 
to the observed Local Void. Not only the galaxy number
counts but also the appearance of the  simulated local void
samples are quite similar to the real local void. This
suggests that, rather than being a crisis for the $\Lambda$CDM framework,
the emptiness of the local void is a triumph of standard $\Lambda$CDM
theory. 

In our simulated local void, the paucity of galaxies is mainly due to
the fact that dark matter haloes are much less abundant than in
the field. The environmental dependence of halo formation caused by
halo assembly bias plays a less important but still sizable role,
which results in $25$ percent lower galaxy abundance in void galaxies 
compared to the field galaxies in dark matter haloes of the same mass.

\section*{acknowledgments}
 We are grateful to I. D. Karachentsev for many very helpful explanations
 of observational data, and to Andrew Cooper, Jie Wang, Ran Li, Lan Wang
 for stimulating discussions. We acknowledge support from NSFC grants 
 (Nos 11143005 and No.11133003 ) and  the Strategic Priority Research 
 Program ``The Emergence of Cosmological Structure''  of the Chinese
 Academy of Sciences (No. XDB09000000).  LG also acknowledges the
 one-hundred-talents program of the Chinese academy of science (CAS),
  the {\small MPG} partner Group family,  and an {\small STFC} Advanced
 Fellowship, as well as the hospitality of the Institute for Computational
 Cosmology at Durham University.  QG also acknowledges a Royal Society
 Newton International Fellowship. 
\bibliographystyle{mnras}
\setlength{\bibhang}{2.0em}
\setlength\labelwidth{0.0em}
\bibliography{void}

\end{document}